\documentclass[iop]{emulateapj}
\newcommand{\oii}{[\ion{O}{2}]}
\newcommand{\oiii}{[\ion{O}{3}]}
\newcommand{\hb}{H$\beta$} 
\newcommand{\ha}{H$\alpha$}
 
\newcommand{\sii}{[\ion{S}{2}]}
\newcommand{\nii}{[\ion{N}{2}]}

\newcommand{\rtt}{$R_{23}$}

\begin{document}
\title{Low Masses and High Redshifts: The Evolution of the Mass-Metallicity Relation\altaffilmark{*}} 
\author{Alaina Henry\altaffilmark{1,5} Claudia Scarlata\altaffilmark{2},  Alberto Dom\'inguez\altaffilmark{3},   Matthew Malkan\altaffilmark{4}, Crystal L. Martin\altaffilmark{5},  Brian Siana\altaffilmark{3}, 
Hakim Atek\altaffilmark{6}, Alejandro G. Bedregal\altaffilmark{2,7}, James W. Colbert\altaffilmark{8}, Marc Rafelski\altaffilmark{8}, Nathaniel Ross\altaffilmark{4}, Harry Teplitz\altaffilmark{9}, 
Andrew J. Bunker\altaffilmark{10}, Alan Dressler\altaffilmark{11}, Nimish Hathi\altaffilmark{11}, Daniel Masters\altaffilmark{3,11},  Patrick McCarthy\altaffilmark{11}, Amber Straughn\altaffilmark{1}} 

\altaffiltext{*}{Based on observations made with the NASA/ESA Hubble Space Telescope, which is operated by the Association of Universities for Research in Astronomy, Inc., under NASA contract NAS 5-26555.}
\altaffiltext{1}{Astrophysics Science Division, Goddard Space Flight Center, Code 665, Greenbelt, MD 20771; alaina.henry@nasa.gov}
\altaffiltext{2}{Minnesota Institute for Astrophysics, University of Minnesota, Minneapolis, MN 55455}
\altaffiltext{3}{Department of Physics and Astronomy, University of California, Riverside, Riverside, CA 92521} 
\altaffiltext{4}{Department of Physics and Astronomy, University of California, Los Angeles, Los Angeles, CA 90095} 
\altaffiltext{5}{Department of Physics, University of California, Santa Barbara, CA 93106}
\altaffiltext{6}{Laboratoire d'astrophysique, \'Ecole Polytechniuqe F\'ed\'erale de Lausanne, Observatoire de Sauverny, 1290, Versoix, Switzerland} 
\altaffiltext{7}{Department of Physics and Astronomy, Tufts University, Medford, MA 02155} 
\altaffiltext{8}{Spitzer Science Center, California Institute of Technology, Pasadena, CA  91125}
\altaffiltext{9}{Infrared Processing and Analysis Center, Caltech, Pasadena, CA 91125}
\altaffiltext{10} {Department of Physics, University of Oxford, Denys Wilkinson Building, Keble Road, OX1 3RH, UK}
\altaffiltext{11}{Observatories of the Carnegie Institution for Science, Pasadena, CA 91101, USA}

\begin{abstract} 
We present the first robust measurement of the high redshift mass-metallicity (MZ) relation at $10^{8}\la M/M_{\sun} \la10^{10}$, 
obtained by
stacking spectra of 83 emission-line galaxies with secure redshifts between 
$1.3\la z\la2.3$.    For these redshifts, infrared grism spectroscopy with the 
 {\it Hubble Space Telescope} Wide Field Camera 3  is sensitive to the \rtt\  metallicity 
 diagnostic: (\oii\ $\lambda\lambda 3726,3729$ + \oiii\ $\lambda \lambda4959,5007$)/\hb.  
   Using spectra stacked in four mass quartiles, we find a MZ relation that declines significantly with 
   decreasing mass, extending from 12+log(O/H) $=8.8$ at $M=10^{9.8}$ M$_{\sun}$, to 12+log(O/H)$=8.2$ at $M=10^{8.2}M_{\sun}$.   
   After correcting for systematic offsets between metallicity indicators, 
 we compare our MZ relation to measurements from  the stacked spectra of galaxies with $M\ga10^{9.5}~M_{\sun}$ and $z\sim 2.3$.  Within the statistical uncertainties, our 
 MZ relation agrees with the $z\sim2.3$ result,  particularly since our somewhat higher metallicities (by around 0.1 dex) are qualitatively consistent with the lower mean redshift 
 ($z=1.76$) of our sample.   For the masses probed by our data, the MZ relation shows a steep slope which is suggestive of feedback from energy-driven winds, and 
 a  cosmological downsizing evolution where high mass galaxies reach the local MZ relation at earlier times.   
 In addition, we show that our sample falls on an extrapolation of the star-forming main sequence (the SFR-M$_{*}$ relation)
 at this redshift. This result indicates that  grism emission-line selected samples do not have preferentially high SFRs. Finally,  we report no evidence for evolution of the  mass-metallicity-SFR plane; our stack-averaged measurements show excellent agreement with the local relation.
\end{abstract}

\section{Introduction} 
 The correlation between stellar mass and gas-phase metallicity in galaxies (the mass-metallicity, or MZ relation) 
 is sensitive to the processes that regulate their growth.
This relation is shaped by gas outflow and accretion rates, the enrichment of these
 gas flows, and the star formation efficiency of galaxies \citep{T04, Brooks, Dalcanton07, Erb08, FD08, PS11, Dave12, Henry13}. 
  It is expected, then, that  observations of the MZ
relation towards low masses and high redshifts will constrain the assembly of galaxies  and the metal enrichment of the intergalactic medium (IGM).

To date, a complete picture of metallicity evolution has not emerged. 
While studies focusing on higher mass galaxies at  $0.5<z<3$  have shown  lower metallicities 
than are seen locally   \citep{Zahid,Shapley05, Erb06, Liu, Maiolino08, Wright,Hayashi},  evolution 
below $M\sim10^{9.0}-10^{9.5}M_{\sun}$ remains unconstrained.   In \cite{Henry13}, we began to probe masses just below 
$M\sim10^{8.5}M_{\sun}$  at intermediate redshifts; nevertheless measuring metallicities 
at higher redshifts ($z>1$) has  been challenging because of the requirement for 
infrared spectroscopy of large, faint samples of galaxies.  
Although lower masses and metallicities have been reached for some strongly lensed galaxies  (e.\ g. \citealt{Teplitz, Hainline09, Wuyts12, Brammer, Yuan, Belli}), the  uncertainties of these measurements are large.

In this Letter,  we present the MZ relation at $z>1.3$,  derived by stacking 83
galaxies with observations covering the \rtt\  metallicity diagnostic, (\oii$\lambda\lambda 3726,3729$ + \oiii$\lambda \lambda4959,5007$)/\hb\ \citep{Pagel}.   Our sample, which is drawn  from  {\it Hubble Space Telescope} Wide Field Camera 3 (WFC3; \citealt{MacKenty}) grism observations, 
reaches $M\sim10^{8}M_{\sun}$ at $1.3<z<2.3$.   
By focusing on measurements from stacked spectra, we avoid bias
introduced by the requirement that \hb\ be detected in each spectrum.
We adopt a \cite{Chabrier} initial mass function (IMF) and take $\Omega_M=0.3$, $\Omega_\Lambda=0.7$ and $H_0=70$ 
km s$^{-1}{\rm Mpc}^{-1}$.  When reporting  measurements of doublet lines (i.e.  \oiii$\lambda \lambda4959,5007$), we use  ``\oii'' or ``\oiii'' to refer to both lines.

\section{Observations and Sample Selection}
 We use
data from  the WFC3 Infrared Spectroscopic Parallel (WISP) Survey, as described by  \cite{Atek10, Atek11}, \cite{D13, Colbert} and \cite{Bedregal}.    The WISP Survey uses both the G102  ($0.8-1.1$\micron, $R\sim210$) and G141 ($1.1-1.7$\micron, $R\sim130$) grisms, enabling measurements of \rtt\ from 
$1.3<z<2.3$.    As the WISP survey is ongoing, we restrict our analysis to  29 fields that have emission-line catalogs and supporting optical imaging, either from WFC3/UVIS (24 fields) or from the Large Format Camera \citep{LFC} on the 200-inch Hale Telescope at Palomar Observatory ($g$ and $i$ bands; 5 fields).  Five-band galaxy spectral energy distributions (SEDs; including {\it Spitzer}/IRAC 3.6 \micron\ photometry for most) are obtained from photometric  catalogs described in \cite{D13} and \cite{Bedregal}.    
 
 Using the WISP emission line catalogs \citep{Colbert},  we identify 128 galaxies with secure redshifts measured from multiple lines.
Objects with contamination from overlapping spectra were removed, leaving 74 galaxies.     Of these, we exclude five galaxies which have inadequate mass constraints:  three fall in the UVIS chip gap, one is detected in only one broad-band filter, and one is blended with another object in ground-based $g$ and $i$ images.   These cuts leave  69 galaxies from WISP.  
 
We supplement this sample with publicly released grism data covering the Ultra Deep Field (UDF)\footnote{http://monoceros.astro.yale.edu/RELEASE\_V3.0/Spectra/UDF
/Web/UDF\_3dhst\_redshift\_v1.0.html};   since the UDF is observed only in G141, \rtt\  is covered only for the 25 galaxies between $2.04<z<2.36$.  Of these,  we exclude three galaxies because of contamination, two because their spectra do not contain emission lines, and one with incomplete imaging near the edge of the UDF.  We construct SEDs for the remaining 19
galaxies using the nine bands of imaging in the HST eXtreme Deep Field (\citealt{Illingworth}).  Photometric catalogs were created by using SExtractor \citep{sextractor} in dual image mode, with F160W as the detection image.  We use MAG\_AUTO to estimate the total flux.   Errors are increased to account for correlated noise in drizzled data \citep{Casertano}.  

Among the 88 galaxies in the combined sample, we identify and remove five AGN candidates (\S \ref{agn}).  
The mean redshift for the remaining 83 galaxies is $z = 1.76$.

 \section{Measurements}

\begin{figure}
\plotone{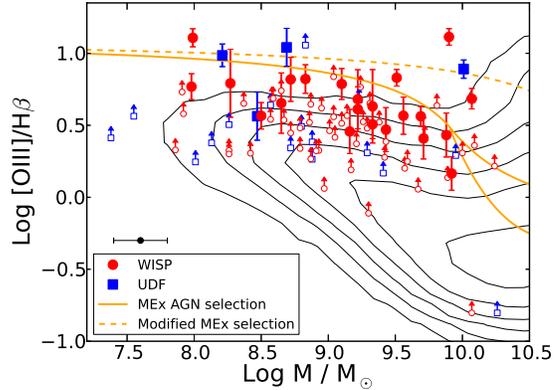}
\caption{The MEx diagrams shows that a minority of WISP and UDF emission line objects are AGN.  Filled symbols show the objects where \hb\ emission 
is detected at $3\sigma$ significance or better (25 galaxies).   Open symbols represent 3$\sigma$ limits (63 galaxies).   The typical stellar mass uncertainty is shown in the lower left.  Contours show the local relation, derived from SDSS DR7 catalogs (http://www.mpa-garching.mpg.de/SDSS).     
Orange lines show AGN thresholds. }  
\label{agn_diagnostics_fig} 
\end{figure}

\subsection{Emission Line Fluxes} 
\label{em_line_meas} 
We measure emission line fluxes from the individual spectra by simultaneously fitting Gaussian profiles  to the  \oii, \oiii, and  \hb lines, as well as  \ha\ and \sii\ when they are covered. Contribution to \ha from unresolved  \nii$\lambda \lambda 6548,6583$ is taken to be negligible, consistent with  our followup  spectroscopy (Masters et al., in prep).   Each region of the spectrum (i.e. \oii, \oiii\ + \hb, \ha + \sii) is also fit with a linear continuum. 
Because the dispersion in the grism spectra diverges from the adopted solution at the red and blue ends, we allow for shifts of up to $\Delta z=\pm 0.01$ between the three widely separated groups of emission lines.   We take the redshift from the fit to \oiii\ and \hb.  Additionally,  since the 
\oiii$\lambda\lambda4959,5007$ doublet is marginally resolved, we fit both lines assuming a 3:1 flux ratio. Finally, the emission lines should be spectrally unresolved with widths  set by the size of the galaxy in the dispersion direction.   Since the dispersion is two times higher in the G102 spectra, we require lines measured in this grism to have widths a factor of  two narrower (in \AA) than those of the redder lines. 
 
\subsection{Stellar Mass} 
\label{mass} 
Stellar masses are derived with the FAST software \citep{FAST},  fitting  BC03 \citep{BC03} template spectra to our broadband photometry.
 We take a grid of stellar population parameters that include: exponentially declining star formation histories with e-folding times ranging from 40 Myr to 10 Gyr;  ages ranging from 50 Myrs to the age of the Universe, and $A_V=0-3$ for a \cite{Calzetti} extinction curve.   Additionally, we use a \cite{Chabrier}  IMF with   metallicities of 0.004, 0.008, and 0.02 ($Z_{\sun}$).  
We removed the emission line contribution to our broadband photometry, estimating the corrections by integrating our continuum subtracted spectra  (see \S \ref{stacking}) under the  bandpasses that cover the line emission.

 \subsection{AGN} 
\label{agn} 
In Figure \ref{agn_diagnostics_fig}, we use the Mass Excitation (MEx) diagram to identify AGN 
that could contaminate our stacked spectra.     This 
diagnostic uses mass as a proxy for the \nii $\lambda 6583$/\ha\  ratio, which is not measured for our sample. 
  The solid orange lines show the AGN threshold from \cite{Juneau}; objects above and to the right are classified as AGN, and objects between the 
solid curves at $M\ga10^{10}M_{\sun}$ are star-forming/AGN composites.    However, the use of the MEx diagnostic is 
compromised by the evolution of the MZ relation.  At fixed metallicity (or \oiii/\hb), galaxies at $z\sim1-2$ are 1-2 dex more massive than $z\sim0$ galaxies 
\citep{Shapley05, Erb06}.  Therefore, we expect that the star-forming locus in Figure 1 shifts to the right at higher redshifts. 
The dashed curve illustrates the possible evolution, by shifting the Juneau et al. curves a conservative 1 dex in stellar mass.   
Taking this modified threshold, we find five AGN candidates.   These objects are excluded from further analysis. 
 We acknowledge that the more aggressive AGN selection would slightly modify our MZ relation above $M>10^{9.4}M_{\sun}$, increasing the metallicity derived in \S \ref{met} by 0.03 dex at $M=10^{9.8}M_{\sun}$.  Nevertheless, this shift falls within our uncertainties, and does not affect the conclusions drawn in \S 4 and \S 5.

For the galaxies that are undetected in \hb, we cannot rule out an AGN  for any given object.  However, given the distribution of detections and limits in Figure \ref{agn_diagnostics_fig}, it is unlikely that a sizable fraction of these galaxies have very high \oiii/\hb\ ratios.   We also note that none of the UDF galaxies are identified as AGN from X-ray counterparts  in the \cite{Xue} catalog.  Therefore, 
we conclude that stacked spectra will be minimally affected by  this contamination.   Overall, the low AGN content in our sample is consistent with the {\it low mass}, grism-selected objects reported by \cite{Trump13}.

\begin{figure} 
\plotone{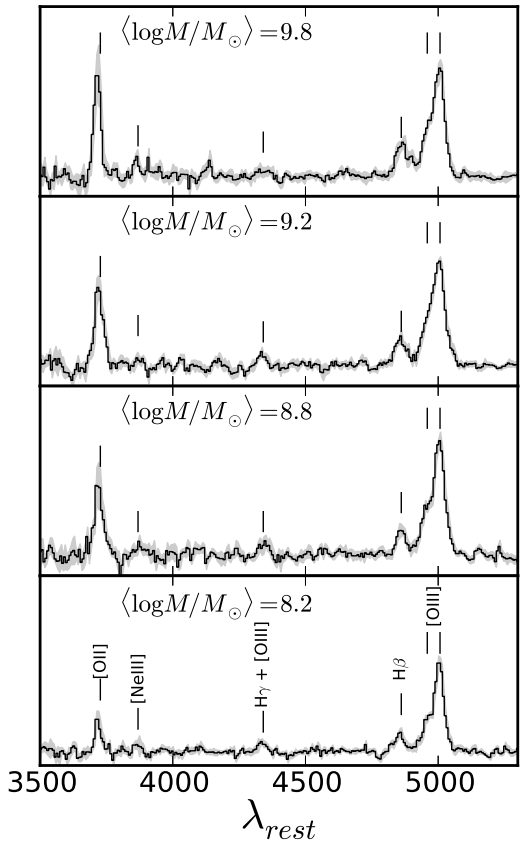} 
\caption{Stacked spectra are shown with shaded regions indicating the   $\pm 1\sigma$ uncertainty. 
Vertical lines mark the expected wavelengths of the line emission, including  H$\gamma $ + \oiii$\lambda 4363$ and [\ion{Ne}{3}]$\lambda3869$. 
We do not interpret these weaker features because of their low S/N and underlying stellar absorption. }  
\label{spec_fig} 
\end{figure}

\subsection{Spectral Stacking} 
\label{stacking} 
In order to stack spectra, we follow a  procedure similar to \cite{D13}.  First, we subtract a model continuum for each galaxy, made by masking the emission lines  and smoothing with a 30 pixel boxcar.  Each spectrum is  then normalized by its \oiii\ flux, so that all galaxies contribute equally.  
 Finally, the spectra are de-redshifted and combined by taking the median. The relative line fluxes are determined by fitting Gaussian profiles.    The measurement follows the same procedure described in \S \ref{em_line_meas}, except the widths of the emission lines are allowed to differ from one another because the relative contribution of the G102 and G141 data varies with rest-wavelength.  Error spectra are generated by bootstrap resampling the spectra in each mass bin to generate 500 artificial stacks, and taking the RMS at each wavelength.  Errors on the relative line fluxes are calculated by propagating the uncertainties on the fit parameters. 

Our sample of 83 galaxies is divided into four mass quartiles 
with $7.3\le{\rm log}~M/M_{\sun}<8.6$, $8.6\le{\rm log}~M/M_{\sun}<9.0$, $9.0\le{\rm log}~ M/M_{\sun}<9.4$, $9.4\le{\rm log}~ M/M_{\sun}<10.2$.  The stacked spectra  are shown in Figure \ref{spec_fig}.

\begin{deluxetable*}{lcccccccccc} 
\tablecolumns{11} 
\tablewidth{0pc} 
\tabletypesize{\footnotesize}
\setlength{\tabcolsep}{0.01in} 
\tablehead{
\colhead{log $M_{\star}/M_{\sun}$}  &  \colhead{N}  &  \colhead{$\langle z \rangle$} & \colhead{EW(\hb)} & \colhead{log \oiii/ \hb} & \colhead{O32} &  \colhead{log \rtt}  & \colhead{12+log(O/H)}     &  \colhead{12+log(O/H)} &  \colhead{SFR}  & \colhead{12+log(O/H)}  \\ 
&  & &  \colhead{ (\AA) } &  & &  & (KK04) & (Maiolino)  & \colhead{(M$_{\odot}$ yr$^{-1}$)} &  \colhead{(FMR)} \\ 
\colhead{(1)}  & \colhead{(2)} & \colhead{(3)} & \colhead{(4)} & \colhead{(5)} & \colhead{(6)} & \colhead{(7)} & \colhead{(8)} & \colhead{(9)} & \colhead{(10)} &\colhead{(11)} 
} 
\startdata 
\cutinhead{Dust Corrected: \ha/\hb = $3.6 \pm 1.1$}
8.2   & 21  & 1.82 &  110   &$0.77 \pm 0.09$  & $0.65 \pm 0.15$  &  $0.85 \pm 0.09$  & $8.16^{ +0.16}_{-0.15}$  & $7.59_{-0.16}^{+0.44}$ & 7  & \nodata  \\
8.8     & 22 & 1.73 &  47  &   $0.81 \pm 0.08$ &  $0.33 \pm 0.14$ &  $ 0.98 \pm 0.08$   & $8.43^{+0.16}_{-0.11}$  & $8.03_{-0.33}^{+0.33}$  & 9 & 8.17  \\
9.2     & 21 & 1.77 &   36 & $0.66 \pm 0.06 $  & $0.24 \pm 0.12$  & $ 0.85 \pm 0.07$  & $8.65^{+0.14}_{-0.11}$  & $8.47_{-0.26}^{+0.14}$  & 13  & 8.33     \\
 9.8 & 19 & 1.74&   29 &  $0.53 \pm 0.07$  & $0.22 \pm 0.13$    & $0.74 \pm 0.08$  & $8.82^{+0.08}_{-0.10}$  &  $8.68_{-0.12}^{+0.09}$ &  21 & 8.60    \\ 
 \cutinhead{No dust correction} 
 8.2  &  21 &1.82  & 110  & $0.76 \pm 0.09$  & $0.76 \pm 0.10$ &  $0.84 \pm 0.09$  & $8.12_{-0.14}^{+0.15} $  &   $7.57_{-0.15}^{+0.37}$   &  3   &     \nodata  \\ 
 8.8  &  22 & 1.73  & 47  & $0.82 \pm 0.08$  & $ 0.45 \pm 0.08 $ &   $0.96 \pm 0.08$  & $8.42_{-0.16}^{+0.30}$   &  $8.03_{-0.38}^{+0.38}$    & 4  &  8.22 \\
9.2  & 21  & 1.77 &   36  & $0.67 \pm 0.06$   & $0.35 \pm 0.05$  &  $0.83 \pm 0.06 $  &  $8.70_{-0.11}^{+0.09} $  &  $8.53_{-0.16}^{+0.11}$   & 6   &  8.39 \\ 
 9.8  &  19 & 1.74  &  29  &  $0.54 \pm 0.07$  & $0.33 \pm 0.08$  & $ 0.71 \pm 0.07 $ & $8.86_{-0.08}^{+0.06}$    & $8.72_{-0.09}^{+0.08}$   & 9    &   8.67 
 \enddata 
\tablecomments{ {\bf (1)}  Mean stellar mass  {\bf (2)}  Number of galaxies in each stack;  {\bf (3)}  mean redshift in each stack;  {\bf  (4)}  Average rest-frame \hb\ equivalent width of the stacked galaxies  (uncorrected for stellar absorption) {\bf  (5)} \oiii/\hb\ flux ratio, corrected for stellar absorption {\bf (6)}  The \oiii/\oii\  ratio (O32 $\equiv$ log \oiii/\oii); {\bf (7)} The metallicity sensitive \rtt\ diagnostic, corrected for stellar absorption; {\bf (8)} Oxygen abundance calculated using the KK04 calibration; {\bf (9)} Oxygen abundance calculated using the \cite{Maiolino08} calibration; {\bf (10)} SFR  estimated from mean \hb\ luminosity of the stacked galaxies; {\bf (11)} The metallicity predicted from the local FMR (\citealt{Mannucci11}; given for the $M\ga10^{8.5}M_{\sun}$ mass bins where the FMR is defined.) } 
\label{results_table} 
\end{deluxetable*}

\subsection{Metallicity} 
\label{met}  
 A few steps must be  taken to infer metallicities from our stacked spectra. 
First, since \oiii\ and \oii\ are widely separated in wavelength,  a dust correction is required.    We estimate 
extinction by measuring the Balmer decrement for  each of the five galaxies in our  $z<1.5$ subsample where both \ha\ and \hb\ are detected (at $3\sigma$ significance or greater).  Consistent with our findings in \cite{D13}, we 
measure an average \ha/\hb\ $=3.6$, with an RMS of 1.1.    We also stack all of the 28 galaxies with $1.3<z<1.5$; here we find \ha/\hb\ $=2.48\pm0.33$, 
consistent with no dust.   In the analysis that follows, we calculate  line ratios assuming both no dust, and  \ha/\hb\ $= 3.6\pm1.1$ (with a \citealt{Calzetti} extinction curve).  
In the latter case, the scatter on the Balmer decrement is included in the metallicity errors. 

Second,  the flux measured from the \hb\  emission line can be artificially reduced by stellar absorption, which is estimated to have an equivalent width (EW) of 3-4 \AA\ in similar grism-selected 
galaxies \citep{D13}.   Since our stacked spectra only represent relative line fluxes,  we estimate the \hb\ EW  from the 
the mean \oiii\ EW of the individual stacked galaxies and \oiii/\hb\ ratio for that stack.  The values for the rest-frame \hb\ EW  range from 30 to 110\AA\ (see Table \ref{results_table}).   The relative \hb\
fluxes are increased by $({\rm EW}_{rest}({\rm H}\beta)+3.5)/{\rm EW}_{rest}$ to account for this absorption. 

\begin{figure} 
\plotone{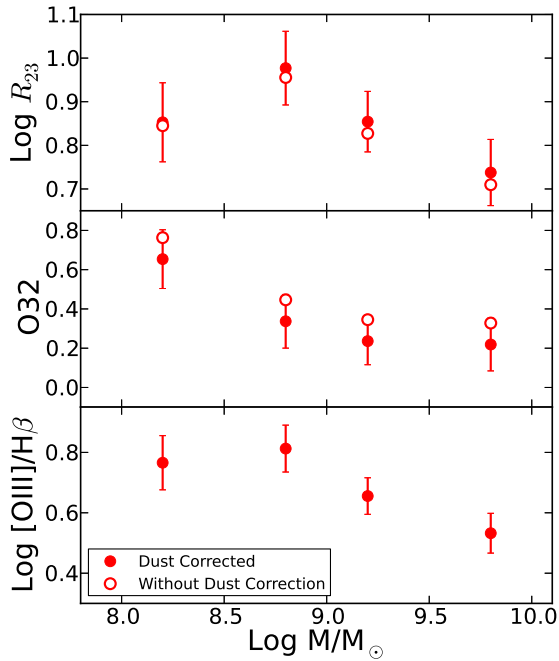}
\caption{Emission line ratios measured from our stacked spectra (with and without dust correction) show that the turn-around metallicity is reached around $M\sim10^{8.8}M_{\sun}$.   
 } 
\label{branch_tests}
\end{figure}

Finally,  \rtt\ is double-valued; a single measurement can correspond to either a high or low metallicity.  Determining whether 
a  galaxy lies on the high or low metallicity branch can be  challenging \citep{KE08, Henry13}.   
Nevertheless,  under the reasonable assumption that metallicity decreases towards lower masses, an average trend is expected: as mass decreases, \rtt\  should increase to log\rtt\  $\sim1$, before turning over and decreasing.   For the first time at high redshift, this trend is observed in Figure \ref{branch_tests}.  We conclude that the $M=10^{8.8}M_{\sun}$ bin marks the ``turn-around," with the two higher mass bins falling on the upper branch of \rtt\ and the lowest mass bin implying lower-branch metallicities.    The middle panel of Figure \ref{branch_tests} supports this conclusion;  as in the local Universe,  the lowest  metallicity galaxies have more highly ionized gas,  as indicated by higher O32 (log \oiii/\oii).   

We have tested that  the results from stacking are robust to scatter between the upper and lower-branch bins.  Taking a simulated MZ relation with the same slope as we have measured, we adopt an intrinsic metallicity scatter of 0.1 dex (as at $z\sim0.1$; \citealt{T04}) and 0.2 dex of scatter in stellar mass  due to our typical measurement error.  We find that fewer than 10\% of the galaxies in the bins adjacent to the turn-around should be attributed to the opposite \rtt\ branch,  and less than 1\% of galaxies in our high-mass bin will fall on the lower branch.   This exercise shows that 
metallicities may be overestimated by 0.03-0.04 dex in the $M=10^{9.2}M_{\sun}$  bin, and underestimated by the same amount in our lowest mass bin.  
We do not apply a correction for this effect, since its amplitude is smaller than our uncertainties. 

Finally, as detailed in \cite{KE08}, different metallicity calibrations of strong-line ratios yield systematically different abundances. 
In Table  \ref{results_table}, metallicities are reported on both the  Kobulnicky \& Kewley (2004; KK04) and \cite{Maiolino08} scale.

\begin{figure*} 
\plottwo{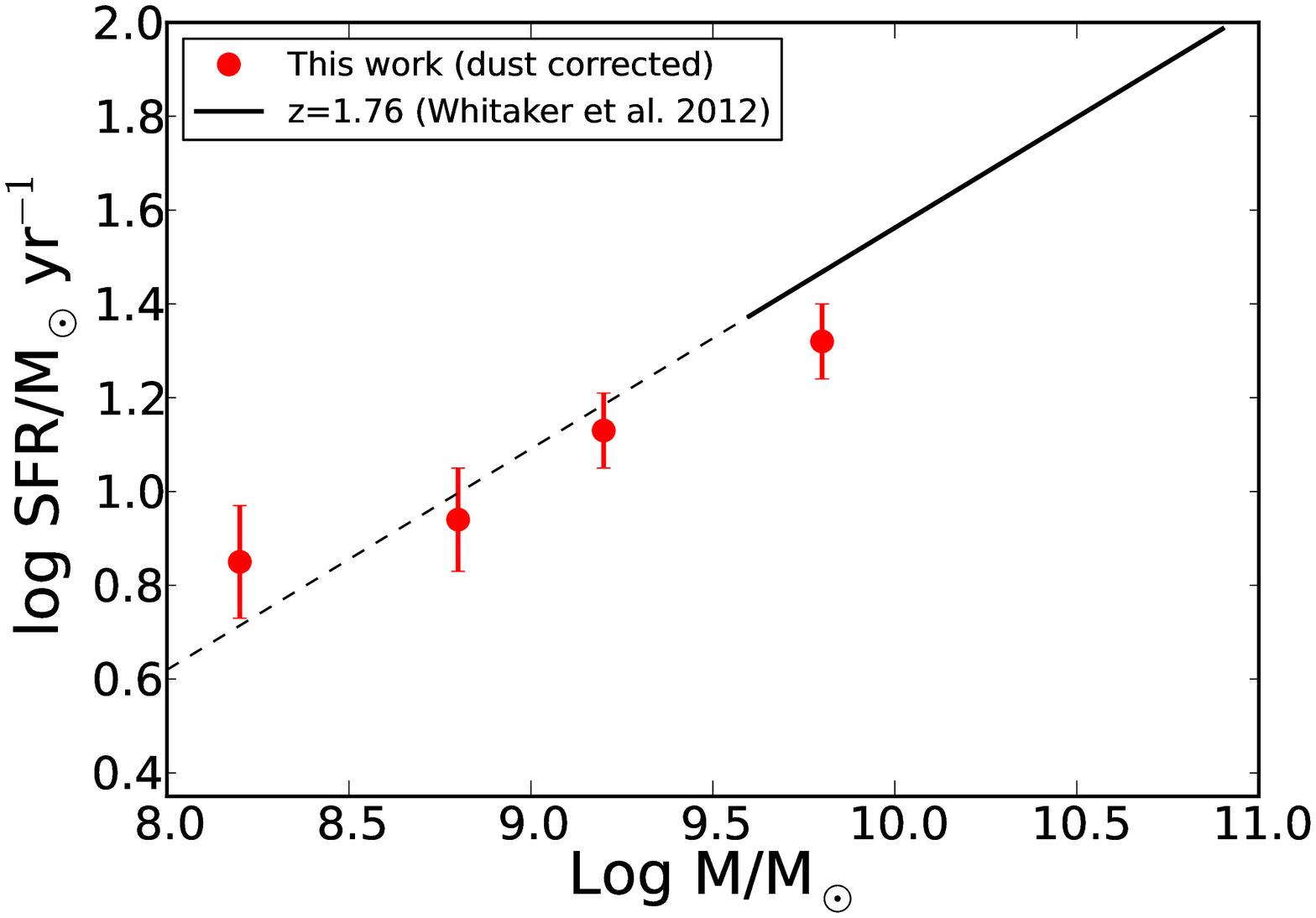} {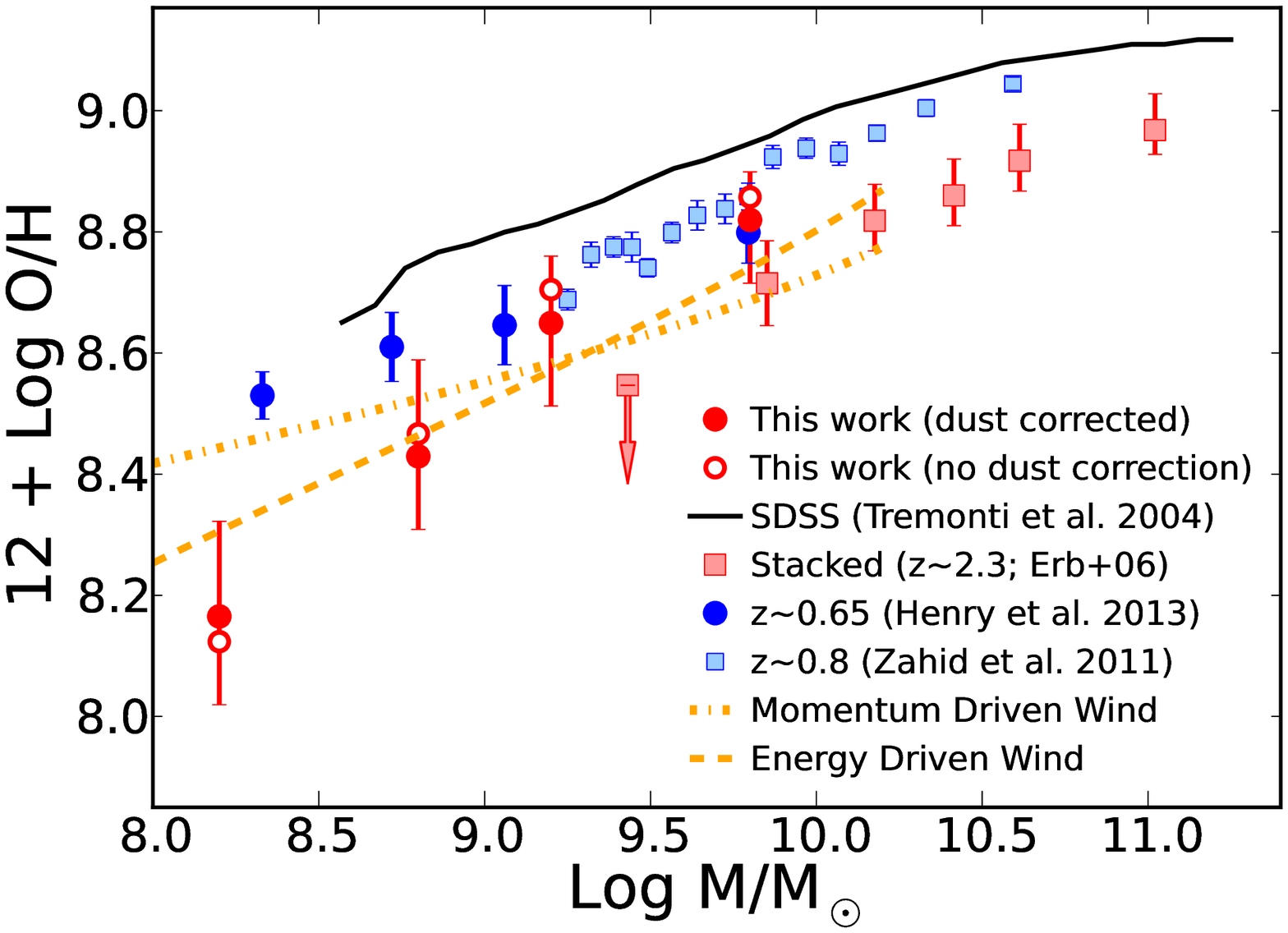} 
\caption{{\it Left:} The dust-corrected SFRs for our stacked spectra fall along an extrapolation of the $z=1.76$ star-forming main sequence, indicated by the dashed line.  Error bars are derived from the error on the mean \hb\ luminosity, and do not include the uncertainties associated with extinction correction.
   {\it Right:}   The MZ relation at $z\sim1.76$ declines steeply towards low stellar masses.   All metallicities are given on the KK04 calibration (when necessary, converted using \citealt{KE08}). 
 Intermediate redshift data from \cite{Zahid}  and \cite{Henry13} are binned averages.   Orange curves (arbitrarily normalized) show how different prescriptions for galactic outflows change the {\it slope}  of the MZ relation.
}
\label{mz_fig}  
\end{figure*}

\section{Star-formation, sample bias, and the fundamental metallicity relation}
In the local Universe, metallicity, stellar mass, and SFR form a plane called the Fundamental Metallicity Relation (FMR; \citealt{Mannucci10}), 
in the sense that galaxies with high specific SFRs have lower metallicities. 
As a result,  it is important to consider selection effects when interpreting the MZ relation.  For example, UV-continuum or emission-line selected samples may
be biased towards higher SFRs and lower metallicities than mass-selected samples.   Therefore, we use \hb\ luminosity 
(measured  analogously to the \hb\ EW, using the mean \oiii\ luminosity and the \oiii/\hb\ ratio) to estimate the SFR for each stack.   
  The \hb\ luminosity is corrected for dust and stellar absorption, and multiplied by 2.86 to obtain the \ha\ luminosity. 
  We then calculate the SFR from the relation in \cite{K98}, converted to a \cite{Chabrier} IMF\footnote{The SFR calibration in \cite{K98} is defined for solar metallicity, and will therefore
overestimate SFRs at low metallicity (by 25\% and 37\% for continuous star-forming, BC03 models with a \citealt{Chabrier} IMF at $Z=0.4Z_{\sun}$  and $Z=0.2Z_{\sun}$).   However, in order to compare to previous measurements of the star-forming main sequence, we report SFRs for solar metallicity.}.
In Table \ref{results_table} we list the average SFR in each mass bin.  
Figure \ref{mz_fig} shows that our dust-corrected SFRs lie very close to an extrapolation of  the $z=1.76$ star-forming main sequence from \cite{Whitaker12}.     
This comparison shows that our grism-selected sample is not likely biased towards high specific SFRs.

Remarkably, \cite{Mannucci10} report that the FMR does not evolve with redshift (at least out to $z\sim3$).    
Further studies confirm this claim for smaller samples of low mass galaxies  \citep{Henry13, Belli}. 
Using our stack-averaged measurements, we can also test for evolution.  In Table 1 we list the  oxygen abundance, predicted from the FMR in \cite{Mannucci11}.  This quantitiy shows excellent agreement with
the metallicities calculated on the \cite{Maiolino08} calibration (on which the FMR is defined). 
We conclude that our sample shows no evidence for evolution of the FMR.

\section{The Mass-Metallicity Relation} 
Figure \ref{mz_fig} shows the MZ relation derived from our stacked spectra, for both the dust corrected and uncorrected measurements.    It 
is clear that the dust correction represents only a small adjustment.  Excluding our
lowest mass bin, our (dust corrected) metallicities are 0.33, 0.17,  and 0.10 dex  lower than the SDSS $z\sim0.1$ measurement at $10^{8.8},10^{9.2}$, and $10^{9.8}M_{\sun}$ \citep{T04}, suggesting more significant evolution in low mass galaxies. A similar trend has been observed in the intermediate-redshift MZ relation (also shown in Figure \ref{mz_fig};  \citealt{Zahid, Henry13}). In fact, in \cite{Henry13} we interpreted this evolution as evidence for downsizing,  where higher mass galaxies evolve onto the local 
MZ relation at earlier times.    While the uncertainty on the evolution measured by the present data remains large, it is consistent with downsizing.   Stacking larger samples of grism spectra can help to reduce the uncertainties, but, at the same time, robust measurements of the local MZ relation are needed for $M\la10^{8.5}M_{\sun}$.    Comparison to local dwarf galaxies remains difficult, as most have metallicities derived directly from electron temperatures (which differ systematically from strong-line metallicities), or suffer from uncertainties regarding the \rtt\ branch  \citep{Berg, Ly}. 

 We also compare our MZ relation to the measurement from $z\sim2.3$ galaxies reported by \cite{Erb06}. 
  These metallicities were measured from  the \nii$\lambda 6583$/ \ha\ ratio in stacked spectra  (using the \citealt{PP04} calibration), and are converted to KK04 metallicities using \cite{KE08}.    Figure \ref{mz_fig} shows that our MZ relation agrees with the results from 
  Erb et al.  The somewhat higher metallicities in our data are consistent with the lower redshift  ($z\sim1.76$) of our sample.  The broad agreement between these measurements suggests that the systematic offsets between metallicity calibrations (which we removed) are not drastically different at low and high redshifts.    
  Finally, while some low mass lensed galaxies have metallicity measurements  at these redshifts (e.g.\ \citealt{Wuyts12}),  large scatter and many weak upper limits preclude 
  meaningful comparisons.

Finally, we  compare our data to theoretical models for the MZ relation.  Following our implementation in 
\cite{Henry13}, we show two models calculated from  the prescriptions in \cite{Dave12}. 
Figure \ref{mz_fig} shows that a model regulated by momentum-driven winds predicts a shallower MZ slope than an energy-driven wind model.   While we showed in \cite{Henry13} that the intermediate redshift MZ relation is reproduced by momentum-driven winds,  the higher redshift relation suggests evolution, as it more closely follows the energy-driven wind model.   It is unlikely that a different choice of metallicity calibration would flatten our observed MZ relation to bring it closer in line with momentum-driven winds;  among the widely used strong-line metallicity calibrations, only \cite{P01} and \cite{P05}  yield MZ relations with significantly shallower slopes compared to a KK04 MZ relation.   However, these shallow slopes may be inaccurate, owing to a saturation of the \oiii\ $\lambda4363$ diagnostic at high metallicities \citep{KE08}.   
Nevertheless, the metallicity errors remain large, so while our data favor energy-driven winds, we cannot rule out the shallower slope associated with
momentum-driven winds.

We have presented the first measurement of the MZ relation from a large sample of  low mass,  high redshift galaxies.   Although emission-line selected, these galaxies are not extreme in their SFRs or metallicities.   Rather, our results show excellent agreement with extrapolations of the star-forming main sequence and MZ relation at similar redshifts.  We conclude that the low-mass, high-redshift MZ relation suggests a downsizing evolution and a preference for feedback from energy-driven winds.

\acknowledgements 
We acknowledge Dawn Erb, Kate Whitaker and Danielle Berg for helpful discussions. 
This research was supported by an appointment to the NASA Postdoctoral Program at the Goddard Space Flight Center, 
administered by Oak Ridge Associated Universities through a contract with NASA.   AH also acknowledges support from HST GO 11696, 12284, and 12568.


\begin{thebibliography} 

\bibitem[Atek et al.(2010)]{Atek10} 
Atek, H., Malkan, M., McCarthy, P.,  et al. 2010, \apj, 723, 104

\bibitem[Atek et al.(2011)]{Atek11} 
Atek, H., Siana, B., Scarlata, C.,   et al. 2011, \apj, 743, 121 



\bibitem[Bedregal et al.(2013)]{Bedregal} 
Bedregal, A., Scarlata, C., Henry, A., et al. 2013, \apj, in press  


\bibitem[Belli et al.(2013)]{Belli} 
Belli, S., Jones, T., Ellis, R.~S., \& Richard, J. 2013, \apj, 722, 141


\bibitem[Berg et al.(2012)]{Berg} 
Berg, D.~A., Skillman, E.~D., Marble, A~R., et al. 2012, \apj,  754, 98




\bibitem[Bertin \& Arnouts(1996)]{sextractor} 
Bertin, E., \& Arnouts, S., 1996, \aap, 117, 393 


\bibitem[Brammer et al.(2012)]{Brammer} 
Brammer, G.~B., S\'anchez-Janssen, R., Labb\'e, I., et al. 2012, \apj, 2012, 758L 


\bibitem[Brooks et al.(2007)]{Brooks}
Brooks, A., Governato, F., Booth, C.~M., et al. 2007, \apj,  655, 17L


\bibitem[Bruzual \& Charlot(2003)]{BC03} 
Bruzual, G. \& Charlot, S. 2003, \mnras, 344, 1000

\bibitem[Calzetti et al.(2000)]{Calzetti}
Calzetti, D., Armus, L., Bohlin, R.~C., et al. 2000, \apj, 533, 682

\bibitem[Casertano et al.(2000)]{Casertano} 
Casertano, S., de Mello, D., Dickinson, M., et al. 2000, \aj, 120, 2747

\bibitem[Chabrier(2003)]{Chabrier} 
Chabrier, G. 2003, \pasp, 115, 763


\bibitem[Colbert et al.(2013)]{Colbert}
Colbert, J.~W., Teplitz, H., Atek, H., et al. 2013, arXiv:1305.1399



\bibitem[Dalcanton(2007)]{Dalcanton07} 
Dalcanton, J. ~J. 2007, \apj, 658, 941 


\bibitem[Dav\'e et al.(2012)]{Dave12} 
Dav\'e, R., Finlator, K., \& Oppenheimer, B.~D., 2012, \mnras, 421, 98 


\bibitem[Dom\'inguez et al.(2013)]{D13}
Dom\'inguez, A., Siana, B., Henry, A., et al. 2013, \apj, 763, 145


\bibitem[Erb et al.(2006)]{Erb06} 
Erb, D.~K.,  Shapley, A.~E., Pettini, M., Steidel, C.~C., Reddy, N.~A., \& Adelberger, K. 2006, \apj, 644, 813

\bibitem[Erb(2008)]{Erb08}
Erb, D.~K. 2008,  \apj,  674, 151



\bibitem[Finlator \& Dav\'e(2008)]{FD08}
Finlator, K. \& Dav\'e, R. 2008, \mnras, 385, 2181 





\bibitem[Hainline et al.(2009)]{Hainline09}
Hainline, K., Shapley, A.~E., Kornei, K.~A. Pettini, M., Buckley-Geer, E., Allam, S., \& Tucker, D.~L. 2009, \apj, 701, 52



\bibitem[Hayashi et al.(2009)]{Hayashi} 
Hayashi, M., Motohara, K., Shimasaku, K.  et al. 2009, \apj, 691, 140



\bibitem[Henry et al.(2013)]{Henry13} 
Henry, A., Martin, C.~L., Finlator, K., \& Dressler, A. 2013, ApJ, 769, 148 



\bibitem[Illingworth et al.(2013)]{Illingworth} 
Illingworth, G.~D., Magee, D., Oesch, P.~A., et al. 2013, arXiv:1305.1931


\bibitem[Juneau et al.(2011)]{Juneau} 
Juneau, S., Dickinson, M., Alexander, D.~M., \& Salim, S. 2011, \apj, 736, 104

\bibitem[Kennicutt(1998)]{K98} 
Kennicutt, R.~C. 1998, \araa, 36, 189


\bibitem[Kewley \& Ellison(2008)]{KE08} 
Kewley, L.~J., \& Ellison, S.~L. 2008, \apj, 681, 1183

\bibitem[Kobulnicky \& Kewley(2004)]{KK04} 
Kobulnicky, H.~A., \& Kewley, L.~J. 2004, \apj, 617, 240 [KK04]


\bibitem[Kriek et al.(2009)]{FAST}
Kriek, M., van Dokkum, P.~G., Labb\'e, I., et al.  2009, \apj, 700, 221


\bibitem[Liu et al.(2008)]{Liu} 
Liu, X., Shapley, A.~E., Coil, A.~L., Brinchmann, J., \& Ma, C.-P. 2008, \apj, 678, 758

\bibitem[Ly et al.(2013)]{Ly}
Ly, C., Malkan, M.~A., Nagao, T., et al. 2013, arXiv:1307.7712

\bibitem[MacKenty et al.(2010)]{MacKenty} 
MacKenty, J.~W., Kimble, R.~A., O'Connell, R.~W., \& Townsend, J.~A. 2010, Proc. SPIE, 7731

\bibitem[Maiolino et al.(2008)]{Maiolino08} 
Maiolino, R., Nagao, T., Grazian, A.,  et al. 2008, \aap, 488, 463

\bibitem[Mannucci et al.(2010)]{Mannucci10} 
Mannucci, F.,  Cresci, G., Maiolino, R., Marconi, A., \& Generucci, A. 2010, \mnras, 408, 2115 

\bibitem[Mannucci et al.(2011)]{Mannucci11} 
Mannucci, F., Salvaterra, R., \& Campisi, M.~A. 2011, \mnras, 414, 1263



\bibitem[Pagel et al.(1979)]{Pagel} 
Pagel, B.~E.~J., Edmunds, M.~G., Blackwell, D.~E., Chun, M.~S, \& Smith, G. 1979, \mnras, 189, 95 


\bibitem[Peeples \& Shankar(2011)]{PS11} 
Peeples, M.~S., \& Shankar, F. 2011, \mnras, 417, 2962 

\bibitem[Pettini \& Pagel(2004)]{PP04} 
Pettini, M., \& Pagel, B.~E.~J. 2004, \mnras, 384, 59 

\bibitem[Pilyugin(2001)]{P01} 
Pilyugin, L.~S. 2001, \aap, 374, 412 

\bibitem[Pilyugin \& Thuan(2005)]{P05}
Pilyugin, L.~S., \& Thuan, T.~X. 2005, \apj, 631, 231 



\bibitem[Shapley et al.(2005)]{Shapley05} 
Shapley, A.~E., Coil, A.~L., Ma, C.-P., \& Bundy, K. 2005, \apj, 635, 1006

\bibitem[Simcoe et al.(2000)]{LFC}
Simcoe, R.~A., Metzger,  M.~R., Small, T.~A.,  \& Araya, G.\ 2000, Bulletin of the American Astronomical Society, 32, 758 



\bibitem[Teplitz et al.(2000)]{Teplitz} 
Teplitz, H.~I., McLean, I.~S., Becklin, E.~E., et al. 2000, \apj, 533, 65

\bibitem[Tremonti et al.(2004)]{T04}
Tremonti,C.~A., et al. 2004, \apj, 613, 898 


\bibitem[Trump et al.(2013)]{Trump13} 
Trump, J.~R., Konidaris, N.~P., Barro, G., et al. 2013, \apj, 763, L6


\bibitem[Whitaker et al.(2012)]{Whitaker12} 
Whitaker, K.~E., van Dokkum, P.~G., Brammer, G., \& Franx, M. 2012, \apj, 754, 29L 

\bibitem[Wright et al.(2009)]{Wright} 
Wright, S.~A., Larkin, J.~E., Law, D.~R. 2009, \apj,, 699, 421

\bibitem[Wuyts et al.(2012)]{Wuyts12} 
Wuyts, E., Rigby, J.~R., Sharon, K., \& Gladders, M. 2012, \apj, 755, 73

\bibitem[Xue et al.(2011)]{Xue} 
Xue, Y.~Q., Luo, B., Brandt, W.~N. , et al. 2011, \apjs, 195, 10


\bibitem[Yuan et al.(2013)]{Yuan} 
Yuan, T.-T., Kewley, L.~J., \& Richard, J. 2013, \apj, 763, 9

\bibitem[Zahid et al.(2011)]{Zahid} 
Zahid, H.~J., Kewley, L.~J., \& Bresolin, F. 2011, \apj, 730, 137





\end{thebibliography}
\end{document}